\documentclass[twocolumn,showpacs,preprintnumbers,amsmath,amssymb,superscriptaddress]{revtex4}

\usepackage{amsmath}
\usepackage{amssymb}

\input epsf
\input rotate

\def\sF{\boldsymbol{\Phi}}
\def\sP{\boldsymbol{\Psi}}
\def\sV{\boldsymbol{\cal V}}
\def\sN{\boldsymbol{\cal N}}
\def\sS{\boldsymbol{\cal S}}
\def\st{\boldsymbol{\tau}}

\def\bea{\begin{eqnarray}}
\def\eea{\end{eqnarray}}
\def\ben{\begin{equation}}
\def\een{\end{equation}}
\def\benu{\begin{enumerate}}
\def\enu{\end{enumerate}}

\def\sss{\scriptscriptstyle\rm}

\def\1var{(\bx_1...\bx\N)}

\def\br{{\bf r}}

\def\bx{{x}}

\def\bq{{\bf q}}

\def\xc{_{\sss XC}}

\def\N{_{\sss N}}
\def\H{_{\sss H}}

\def\sph_int{ {\int d^3 r}}

\def\infintd3r{ \int_{-\infty}^\infty d^3r\,}
\def\intd3r{ \int d^3r\,}

\def\laplace1d{\frac{d^2}{dx^2}}
\def\plaplace1d{\frac{d^2}{d{x'}^2}}

\def\padr2{\frac{\partial^2}{\partial r^2}}

\begin{document}

\title{{$N$-Representability and stationarity in time-dependent density functional theory}}
\author{Morrel H. Cohen}
\affiliation{Department of Physics and Astronomy, Rutgers University, 126 Frelinghuysen Rd., Piscataway, NJ 08854-8019, USA}
\affiliation{Department of Chemistry, Princeton University, Washington Rd., Princeton, NJ 08544-1009, USA}
\author{Adam Wasserman}
\affiliation{Department of Chemistry and Chemical Biology, Rutgers University, 610 Taylor Rd., Piscataway, NJ 08854-8087, USA}

\date{\today}
\begin{abstract}

To construct an $N$-representable time-dependent
density-functional theory, a generalization to the time domain of
the Levy-Lieb (LL) constrained search algorithm is required. That
the action is only stationary in the Dirac-Frenkel variational
principle eliminates the possibility of basing the search on the
action itself. Instead, we use the norm of the partial functional
derivative of the action in the Hilbert space of the wave
functions in place of the energy of the LL search. The electron
densities entering the formalism are $N$-representable, and the
resulting universal action functional has a unique stationary
point in the density at that corresponding to the solution of the
Schr\"{o}dinger equation. The original Runge-Gross (RG)
formulation is subsumed within the new formalism. Concerns in the
literature about the meaning of the functional derivatives and the
internal consistency of the RG formulation are allayed by
clarifying the nature of the functional derivatives entering the
formalism.

\end{abstract}

\maketitle

\begin{section}{Introduction}

Density-functional theory (DFT) now provides the conceptual,
theoretical, and computational framework for the study of the
ground-state properties of a vast array of quantum-mechanical
systems at all levels of aggregation from atomic to macroscopic.
The foundations for the contemporary theory of chemical reactivity
emerge naturally from DFT as well \cite{PY89,GPL03,CW03}. The
essential elements of DFT are the Hohenberg-Kohn (HK) theorems
\cite{HK64}, the Kohn-Sham (KS) equations \cite{KS65}, the
Levy-Lieb (LL) constrained search algorithm \cite{L79,L82} which
together with the Harriman-Zumbach-Masche (HZM) construction
\cite{H80} introduces $N$-representable densities into DFT;
accurate approximate functionals \cite{PBE96}; and powerful
computational algorithms \cite{CP85}.

As defined through the LL algorithm, the density functional $E[n]$
has a unique global minimum at the ground-state density within the
space $\sN$ of all allowable \cite{L82} electron densities
$n(\br)$. This variational principle of DFT stands in one-to-one
correspondence with the Rayleigh-Ritz variational principle for
the time-independent Schr\"{o}dinger equation and provides the
same generality to the derivation of the KS equations.

Following the ground-breaking HK paper, a series of steps was
taken towards a time-dependent density-functional theory (TDDFT)
\cite{12,GUG95} which culminated in a more general formulation by
Runge and Gross (RG) \cite{RG84}. TDDFT is now being routinely
applied to the calculation of excitation energies of atoms and
molecules \cite{C95}, as well as various physical properties
within the linear response regime \cite{ZS80} and beyond it
\cite{DBG97} (see ref.\cite{MBAG01} for a survey of recent
applications).

In parallel to this success, discussions regarding the foundations of
the theory continue to take place \cite{MBAG01}. In their original
work, Runge and Gross \cite{RG84} employed the quantum-mechanical
action integral (hereon the RG action) to derive time-dependent
Kohn-Sham equations through the Dirac-Frenkel variational principle
\cite{15}. It was later argued \cite{GDP96} that the RG action led to
paradoxes when calculating response functions because these must be
causal, whereas second functional derivatives of the RG action were
thought to be symmetric. This ``symmetry-causality paradox'' was
resolved first by Rajagopal \cite{R96}, who introduced an action based
on the time path introduced by Jackiw and Kerman \cite{JK79}, and
subsequently by van Leeuwen \cite{L98}, who reformulated TDDFT
replacing the RG action by a Keldysh action \cite{K65}. The RG,
Jackiw-Kerman, and Keldysh actions are defined \emph{only} for
time-dependent $v$-representable densities (TDVR), and, regarded as
functionals \emph{only} of the density, are not stationary at the
density of the solution of the time-dependent Schr\"{o}dinger equation
\cite{GUG95,GDP96,BG98,L98,L01}. Such a lack of stationarity is a
decided inconvenience, but even within TDVR TDDFT stationarity can be
restored by recognizing that the density and external potential can be
treated as independent functions \cite{GD88}. Nevertheless, for
reasons analogous to those applying to the ground-state theory, it is
important to generalize the definition of the action functional to
hold for time-dependent $N$-representable densities (TDNR). More
explicitly, Mearns and Kohn \cite{MK87} have shown that small,
time-dependent additions to the ground-state density need not be
$v$-representable in first order. A suitable generalization can be
effected by constructing a constrained-search algorithm for TDDFT
analogous to the LL algorithm for DFT.

Apart from restoring stationarity to the action in TDDFT,
$N$-representability is important because, as in DFT, accurate
solution of the KS equations requires iteration to
self-consistency. The most convenient starting densities may well not
be $v$-representable, nor may the densities be at intermediate stages
of the computational algorithms. It is then essential to have an
action functional and KS potentials defined for $N$-representable
densities both as a matter of principle and for practical reasons.

In this paper we formulate an $N$-representable TDDFT based on
the Dirac-Frenkel variational principle in which the RG action
functional is stationary with respect to $n$ at that unique $n$
derivable from the solution of the time-dependent Schr\"{o}dinger
equation. We establish a one-to-one invertible map between all
densities in a time-dependent generalization of $\sN$ and wave
functions by use of the norm of the partial functional
derivative \cite{BB82} of the RG action in the Hilbert space of the wave
functions. Insertion of that map into the action defines the
action functional. The Runge-Gross formulation of TDVR is subsumed
within this TDNR TDDFT, and the desired stationarity and
generality are achieved.

In Section II, we begin by reviewing two topics central to our later
developments, the Dirac-Frenkel variational principle and the action
and its total and partial derivatives. Via Section II we introduce our
notation for wave functions, operators, functional derivatives,
Hilbert spaces, and more general function spaces. We also introduce
the notion of mapping between abstract spaces as central to the
formulation of TDDFT, following Dreizler and Gross for DFT \cite{DG90}.
In Section III, we recapitulate the RG formulation of $v$-representable TDDFT and
show explicitly that its unnecessary limitation to the density generated by that $v$ which enters the Hamiltonian
destroys the stationarity of the action functional. We also
provide an explicit explanation of why there are no
inconsistencies in the functional derivatives entering the theory
and why second functional derivatives of the RG action with
respect to the density are not symmetric. Up to this point, our paper has concerned itself with the clarification of existing work on $v$-representable TDDFT. In Section IV, we turn to the problem of establishing a satisfactory $N$-representable TDDFT. We begin by stating a set of criteria that such a theory must meet. Next, we review existing proposals (\cite{GD88},\cite{KD86}) and show that they do not meet all of the criteria. Finally, we develop the principal result of this paper, a constrained search algorithm which meets all of the criteria.
We close with a brief summary of our results in Section V.

\end{section}

\begin{section}{Background and notation}

\begin{subsubsection}*{The Dirac-Frenkel variational principle}
Consider a finite system of electrons and nuclei containing $N$
electrons. Ignoring nuclear kinetic energy, keeping the nuclei
fixed, and discarding the internuclear interaction energy as an
irrelevant constant, the system Hamiltonian becomes \ben \hat{\cal
H}[v]=\hat{T}+\hat{W}+\hat{V}[v]=\hat{H}+\hat{V}[v]. \label{eq1}
\een In Eq.(\ref{eq1}), $\hat{T}$ is the electron kinetic-energy
operator and $\hat{W}$ the electron-electron interaction operator.
The operator $\hat{V}[v]$ is the energy of interaction of the
electrons with a time-dependent external potential  $v(\br,t)$,
\ben \hat{V}[v]=\int{d\br~{v(\br,t)\hat{n}(\br)}}~~. \label{eq2}
\een In Eq.(\ref{eq2}), $\hat{n}(\br)$ is the electron-density
operator. $v(\br,t)$ is comprised of the potential energy of an
electron in the fixed nuclear electrostatic potential plus that in
a time-dependent potential generated by sources external to the
system. For each time $t$ in the interval $(t_0,t_1)$ under
consideration, the $\br$-dependence of $v(\br,t)$ must meet the
conditions imposed by Lieb \cite{L82}. In addition, we impose the
requirement that \ben
v(\br,t)\rightarrow{0}~~,~~r\uparrow\infty~~,~~\forall~t~\in(t_0,t_1)
\label{eq3} \een to eliminate irrelevant phase factors in the wave
functions (see also ref.\cite{DG87}). The time dependence of $v$
must meet certain implicit integrability conditions discussed
below. Such acceptable potentials lie in the space $\sV$.
The space ${\mathbb R}^3\times(t_0,t_1)$ is the support on which
the elements $v$ of ${\sV}$ are defined. As indicated by our
notation, $\hat{V}[v]$ is a linear functional of $v$,
Eq.(\ref{eq2}), and so, consequently, is $\hat{\cal H}[v]$,
Eq.(\ref{eq1}).

The wave-functions $\Phi(t)$ of the $N$-electron system are
time-dependent, normalized, antisymmetric functions of the $N$
space and spin coordinates of the electrons, \ben
\left|\left|\Phi(t)\right|\right|=\left(\Phi(t),\Phi(t)\right)=1~~,~~\forall~{t}\in(t_0,t_1)~~.
\label{eq4} \een They satisfy the time-dependent Schr\"{o}dinger
equation (atomic units are used throughout), \ben i\partial_t\Phi(t)=\hat{\cal H}[v]\Phi(t)~~.
\label{eq5} \een Once the initial condition \ben \Phi(t_0)=\Phi_0
\label{eq6} \een is imposed, $\Phi(t)$ is unique, \ben
\Phi(t)={\cal T}_L\exp\left[{-i\int_{t_0}^{t}dt'\hat{\cal
H}[v]}\right]\Phi_0~~. \label{eq7} \een In Eq.(\ref{eq7}) ${\cal
T}_L$ is the time-ordering operator, later to the left.
Eq.(\ref{eq7}) defines implicitly the conditions which $v(t)$,
$\Phi(t)$, and $\Phi_0$ must meet. In addition to those conditions
which were specified by Lieb \cite{L82}, $\Phi(t)$ must be differentiable in
time. The set of such functions which are solutions of
Eq.(\ref{eq5}) for all $v$ in ${\sV}$ form a Hilbert space $\sF$.
They are supported in $\sF$ on the space $\st$, which is the
product of $(t_0,t_1)$ with the configuration and spin space $\sS$
of the $N$-electrons, \ben \st={\sS}\times(t_0,t_1)~~. \label{eq8}
\een All scalar products like the norm entering Eq.(\ref{eq4}) are
defined on ${\sS}$.

Eqs.(\ref{eq5}) plus (\ref{eq6}) implicitly, and (\ref{eq7})
explicitly, define a mapping $M_1:{\sV}\rightarrow\sF$. $M_1$ is
surjective; $\sF$ contains no element which is not associated with
an element of ${\sV}$ \cite{21}. That $M_1$ is injective as well,
i.e. one-to-one and therefore bijective or invertible, can be seen
as follows. Suppose there is a $v'$ and therefore a $\hat{V}'$
which yields the same $\Phi$ as solution of Eq.(\ref{eq5}) as does
$v$ and $\hat{V}$. Subtracting the two Schr\"{o}dinger equations
leads to \ben
\left(\hat{V}'-\hat{V}\right)\Phi=\int{d\br}\left[v'(\br,t)-v(\br,t)\right]\hat{n}(\br)\Phi(t)=0~~,
\label{eq9} \een
which implies that $v'$ must equal $v$ under the
conditions on the $\br$-dependence of $v$ required for the
analogous proof for the time-independent problem (cf
ref.\cite{L82} and p.5 of ref.\cite{DG90}). Thus $M_1^{-1}$ exists
and $\sF\leftrightarrow{\sV}$ is one-to-one. $\Phi$ can then be
regarded as a functional of $v$, $\Phi[v]$, or $v$ one of $\Phi$,
$v[\Phi]$.

Let us now expand the Hilbert space $\sF$ to $\sP$ which contains
all functions $\Psi$ which meet the conditions imposed on $\Phi$
including $\Psi(t_0)=\Phi_0$, except that the $\Psi$ need not
satisfy Eq.(\ref{eq5}). The Dirac-Frenkel variational principle
states that $\Psi$ satisfies Eq.(\ref{eq5}) if and only if
\begin{subequations}
\ben \left(\delta\Psi,\left[i\partial_t-\hat{\cal
H}[v]\right]\Psi\right)=0~~, \een \ben
\left(\delta\Psi(t),\Psi(t)\right)=0~~,~~\forall
t~~.\label{DiracFrenkel}\een\end{subequations}
\end{subsubsection}
\begin{subsubsection}*{The action and its total and partial functional derivatives}

We can now define
the usual quantum-mechanical action functional $A[\Psi,v]$ on the
space $\sP\times{\sV}$, \ben
A[\Psi,v]=\int_{t_0}^{t_1}dt\left(\Psi,\left[i\partial_t-\hat{\cal
H}[v]\right]\Psi\right)~~. \label{eq10} \een $A[\Psi,v]$ is
stationary only at $\Phi[v]$ in $\sP$ with respect to variations
$\delta\Psi$, $\delta\Psi^*$ taken at {\em constant} $v$, given that
$\Psi(t_0)=\Phi_0~\forall\Psi\in\sP$, and requiring as well that
\cite{22} \ben \left(\delta\Psi(t_1),\Psi(t_1)\right)=0~~,
\label{eq11} \een a less severe restriction than that of
Eq.(\ref{DiracFrenkel}). The functional gradient of $A[\Psi,v]$ along
$\Psi^*$ ($v$ is fixed),
\ben
\Theta_{\Psi^*}=\partial_{\Psi^*}A[\Psi,v]=\left[i\partial_t-\hat{\cal
H}[v]\right]\Psi~~, \label{eq12} \een thus vanishes in $\sP$ at
$\Phi[v]$, yielding the time-dependent Schr\"{o}dinger equation.  Note
the use in Eq.(\ref{eq12}) of $\partial_{\Psi^*}$ as a symbol for a
partial functional derivative. We now clarify the nature of such a
derivative. The action is a functional of two functions defined in two different spaces. Accordingly, it does not fit simple
examples of functionals used to define Fr\'{e}chet and G\^{a}teaux
derivatives \cite{BB82}-\cite{LS03}, which restrict the functions on
which they are defined to a single Banach space (in the case of a
Fr\'{e}chet derivative), or normed space (in the case of a G\^{a}teaux
derivative). The G\^{a}teaux derivative has been regarded as a generalization of the concept
of the partial derivative of a function \cite{BB82}. Similarly, the Fr\'{e}chet derivative has been
regarded as a generalization of a total derivative \cite{BB82}. In our case, the
properties of the action are such that taking derivatives only with respect
to $n$ meets the criteria for a Fr\'{e}chet derivative despite the fact
that it is a partial functional derivative. In the following we shall
use the terminology {\em partial functional derivative} to refer to
derivatives with respect to a single function of functionals of more than one function. When, however, we map the potential in the action
back to the density or vice versa so that the action becomes a
functional only of a single function, we shall refer to the functional
derivative taken with respect to that single function as a {\em total
functional derivative}. The total differential could then be
represented as a linear combination of partial functional derivatives
times the corresponding differentials of the respective functions.
\end{subsubsection}
\end{section}

\begin{section}{v-representable TDDFT}

\begin{subsubsection}{$v$-representability and stationarity of the RG action}
\label{sec:3.1} We can recast the arguments of RG\cite{RG84} as
follows. Restrict the argument $\Psi$ of $A[\Psi,v]$ in
Eq.(\ref{eq10}) to lie in $\sF$, the space of $v$-representable
wave functions $\Phi[v']$, defining an action functional, \ben
A[\Phi,v]=\int_{t_0}^{t_1}dt\left(\Phi,\left[i\partial_t-\hat{\cal
H}[v]\right]\Phi\right) \label{eq13p} \een on the space
$\sF\times\sV$. Stationarity of $A[\Psi,v]$ implies stationarity
of $A[\Phi,v]$, i.e. that its partial functional derivative
vanishes, \ben
\partial_{\Phi^*}A[\Phi,v]=0~~.
\label{eq14p}
\een
since $\sF\subset\sP$ and the stationary point of $A[\Psi,v]$ is in $\sF$.

$A[\Phi,v]$ can be established as a functional of $v$ alone by
inserting in $A[\Phi,v]$ that $\Phi[v']$ for which $v'=v$, \ben
A[v]=A[\Phi[v],v]~~. \label{eq15p} \een The stationarity condition
(\ref{eq14p}) then implies that the total functional derivative of
$A[v]$ is $-n$, \ben
\delta_{v(\br,t)}A[v]=\delta_{v(\br,t)}A[\Phi[v],v]=-n(\br,t)~~,
\label{eq16p} \een a generalization of the Hellmann-Feynman
theorem \cite{H33}. The total functional derivative (\ref{eq16p}) does
not vanish, obviously. It is only the partial functional derivative
(\ref{eq14p}) which yields stationarity \cite{GD88}.

To go on to the density-functional, $A[n,v]$, requires
establishing that the map $M_2':\sF\rightarrow{\sN}_v$, \ben
n(\br,t)=\left(\Phi(t),\hat{n}(\br)\Phi(t)\right)~~, \label{eq13}
\een is one-to-one and invertible. In Eq.(\ref{eq13}), $n(\br,t)$
is the time-dependent electron density, and the symbol ${\sN}_v$
stands for the subset of all such $v$-representable densities
contained in ${\sN}$, the time-dependent generalization of the
space of densities of DFT \cite{L82}. All $n(\br,t)$ in ${\sN}$
and ${\sN}_v$ obey the initial condition \ben
n(\br,t_0)=n_0(\br)=\left(\Phi_0,\hat{n}(\br)\Phi_0\right)~~.
\label{eq14} \een Eq.(\ref{eq13}) defines what is meant by the
phrase TDVR; a TDVR density is derivable via Eq.(\ref{eq13}) from
the solution $\Phi$ of the Schr\"{o}dinger equation (\ref{eq5})
for some $v$ in ${\sV}$. Demonstrating the invertibility of $M_2'$
directly, however, is nontrivial. The HZM construction \cite{H80}
shows that $\sP\rightarrow{\sN}$ is many to one.

RG followed an alternative path. Substituting Eq.(\ref{eq7}) into
Eq.(\ref{eq13}) defines a map
$M_3=M_1M_2':{\sV}\rightarrow{\sN}_v$. They then show by a pretty
argument that $M_3$ is one-to-one and invertible for all
potentials $v(t)$ which possess a Taylor expansion in time about
$t_0$ converging for all $t\in(t_0,t_1)$. Van Leeuwen \cite{L01}
has pointed out that it is sufficient for a Taylor series to exist
about a set of points $t_i\in(t_0,t_1)$ for which the radii of
convergence overlap to cover $(t_0,t_1)$.
$M_2'^{-1}:\sN_v\rightarrow\sF$ can then be constructed as
$M_3^{-1}M_1$. Substitution of $M_2'^{-1}$, that is $\Phi[n]$ into
$A[\Phi,v]$, then yields the desired functional $A[n,v]$. $A[n,v]$
is stationary with respect to variation of $n$ at fixed $v$
\cite{GD88}, that is its partial functional derivative vanishes, \ben
\partial_n{A[n,v]}=0~~.
\label{eq19p} \een
It is important to recognize that $A[n,v]$ is
defined for all $n$ generated via Eqs.(\ref{eq7}) and (\ref{eq13})
from some $v'$, which can be varied independently of $v$. It is
only at the stationary point that $v'=v$.

\end{subsubsection}
\begin{subsubsection}{Time-dependent Kohn-Sham equations}
\label{sec:3.2} Substitution of both $M_2'^{-1}$ and $M_3^{-1}$,
i.e. $\Phi[n]$ and $v[n]$, into $A[\Phi,v]$, then yields a
functional $A[n]$ of $n$ only (for a given initial state
\cite{MB01}), the RG action functional. One thus has the option of
using $n$ or $v$ as the independent variable in the functional.
Van Leeuwen \cite{L01} gives a simple and elegant argument for the
construction of the TDKS equations from $A[n]$ without invoking
stationarity in $\sN_v$. Switching now to $A[n]$ from $A[v]$, we
carry out a Legendre transformation to \ben
B[n]=A[n]+\int_{t_0}^{t_1}dt\int{d\br}~v(\br,t;[n])n(\br,t)~~.
\label{eq16} \een From Eq.(\ref{eq16}), it follows that \ben
\delta_{n(\br,t)}B[n]=v(\br,t). \label{eq17} \een The TDKS
equations \cite{L01} follow from (\ref{eq17}).

Consider a system of non-interacting electrons denoted by
subscript $s$ which move in an external potential $v_s(\br,t)$,
starting from a single determinantal state $\Phi_{0s}$ at $t_0$.
$v_s$ is a functional of their electron density, $v_s[n_s]$. The
HZM construction \cite{H80} allows identification of $n_s(\br,t)$
with the density of the interacting system, \ben
n_s(\br,t)\equiv{n(\br,t)}~~,~~\forall\br,t\in(t_0,t_1)~~.
\label{eq18} \een Thus $v_s$ can be regarded as a functional of
$n$. Combining \ben {\delta_{n(\br,t)}{B_s[n]}}=v_s(\br,t)
\label{eq19} \een with Eq.(\ref{eq17}) leads to \ben
v_s(\br,t)=v(\br,t)-{\delta_{n(\br,t)}(A-A_s)}~~. \label{eq20}
\een The usual rearrangements in $A-A_s$ in turn lead to \ben
v_s(\br,t)=v(\br,t)+v\H(\br,t)+v\xc(\br,t)~~, \label{eq21} \een
\ben v\H(\br,t)=\int{d\br'\frac{n(\br',t)}{|\br-\br'|}}~~,
\label{eq22} \een \ben v\xc(\br,t)=-\delta_{n(\br,t)}{A}\xc[n]~~,
\label{eq23} \een
\begin{eqnarray}
A\xc[n]=\int_{t_0}^{t_1}dt\left\{\left[\left(\Psi,i\partial_t\Psi\right)-\left(\Psi_s,i\partial_t\Psi_s\right)\right]\right.\\
\left.-\left[(T-T_s)-(W-W\H)\right]\right\}~~. \label{eq24}
\end{eqnarray} $T$ is the kinetic energy and $W$ the energy of
electron-electron interaction of the interacting electrons in
state $\Phi[n]$. $T_s$ is the kinetic energy of the noninteracting
electrons in state $\Phi_s[n]$. $W\H$ is the Hartree approximation
to $W$ using $\Phi[n]$ or equivalently $\Phi_s[n]$.

It is at this point that concern about the meaning of the functional
derivative defining $v\xc$, Eq.(\ref{eq23}), arises in the literature
\cite{GUG95,GDP96,BG98,L98,L01}. Since in Section IV we shall base
our development of $N$-representable TDDFT on the RG action and since the
above-mentioned concern raises doubts about the validity of doing
this, we now summarize the debate and show why the RG action is
perfectly suitable for the developments of Section IV.
\end{subsubsection}

\begin{subsubsection}{The symmetry-causality dilemma}
\label{sec:3.3} Taking the functional derivative of
Eq.(\ref{eq21}) with respect to $n$ results in \cite{GK85}\ben
\chi^{-1}(\br,t;\br',t')=\chi_s^{-1}(\br,t;\br',t')+f(\br,t;\br',t')~~,
\label{eq25} \een where \ben
\chi(\br,t;\br',t')=-\frac{\delta{n(\br,t)}}{\delta{v}(\br',t')}~~,
\label{eq26} \een \ben
\chi_s(\br,t;\br',t')=-\frac{\delta{n}(\br,t)}{\delta{v_s}(\br',t')}~~,
\label{eq27} \een \ben
f(\br,t;\br',t')=\frac{\delta[v\H+v\xc](\br,t)}{\delta{n}(\br',t')}=\frac{\delta(t-t')}{|\br-\br'|}+\frac{\delta{v\xc}(\br,t)}{\delta{n}(\br',t')}~~.
\label{eq28} \een From Eqs.(\ref{eq13}) and (\ref{eq7}), the well
known retarded character of the time dependence of the
susceptibilities $\chi$ and $\chi_s$ follows; they vanish if
$t'>t$. Their inverses $\chi^{-1}(\br,t;\br',t')$ and
$\chi_s^{-1}(\br,t;\br',t')$ entering Eq.(\ref{eq25}) are retarded
as well. Yet \ben
f\xc(\br,t;\br',t')=\frac{\delta{v}\xc(\br,t)}{\delta{n}(\br',t')}=-\frac{\delta^2A\xc[n]}{\delta{n}(\br',t')\delta{n}(\br,t)}
\label{eq29} \een is formally a second derivative of $A\xc[n]$
according to Eq.(\ref{eq23}). Van Leeuwen \cite{L98,L01} assumes
that, as $f\xc$ is a second functional derivative, it must be
symmetric in $\br,t$ and $\br',t'$. Such symmetry is inconsistent
with the retarded nature of $\chi^{-1}$ and $\chi_s^{-1}$ in
Eq.(\ref{eq25}). Van Leeuwen \cite{L98,L01} describes this
inconsistency as a ``paradox'' and develops TDVR TDDFT from the
Keldysh action instead of the RG action to avoid it. The second
functional derivatives remain symmetric on the Keldysh time
contour but become retarded when mapped into real time.

Gross, Dobson, and Petersilka \cite{GDP96}, on the other hand,
suppose that Eq.(\ref{eq25}) holds and that $f\xc(\br,t;\br',t')$
must be retarded and not symmetric in $\br,t$ and $\br',t'$. They
then conclude by supposing from Schwarz's lemma \cite{W71} that
(1) $f\xc(\br,t;\br',t')$ cannot be a second functional derivative
and that (2) the exact $v\xc[n]$ cannot therefore be a functional
derivative. They conclude further that this in turn is in
contradiction to the principle of stationary action which leads to
$v\xc$ as a functional derivative.

To complicate matters further, Harbola and Banerjee \cite{HB99} have argued that there is no symmetry-causality dilemma because while $\chi$ is causal, $\chi^{-1}$ is symmetric. Amusia and Shaginyan \cite{AS01}, while not disagreeing with this conclusion, have argued that, in contrast, it is possible to construct a causal $\chi^{-1}$ as well. Harbola \cite{H01} has responded that reference \cite{AS01} itself implies a causality in the potential as a functional of the density. van Leeuwen, however, has argued that $\chi^{-1}$ must be rigorously causal in analogy with the properties of discrete lower triangular matrices \cite{L01}, an argument which does not take into account the fact that $\chi^{-1}$ is not a smooth function of $t-t'$ but contains both a delta function and the second derivative of a delta function at $t=t'^+$. We show in Appendix A that $\chi^{-1}(t-t')$ is causal, consisting of those singular functions at $t=t'^+$ plus a smooth causal function of $(t-t')$, so that the dilemma remains.

\end{subsubsection}
\begin{subsubsection}{A way out of the dilemma}
\label{sec:3.4} We conclude that in the context of TDDFT at the RG
level, $\chi^{-1}$ is causal. The most forceful argument that the
causality of $\chi^{-1}$ imposes a symmetry-causality dilemma via
Eq.(\ref{eq25}) is that of Gross, Dobson, and Petersilka
\cite{GDP96}. The flaw in their reasoning is the supposition that
Schwarz's lemma can be applied to the functionals of TDDFT.
Throughout all of density-functional theory, the Fr\'echet
definition \cite{BB82} of the functional derivative was implicitly
used. In the present instance, the functional derivatives of
Eqs.(\ref{eq19p}),(\ref{eq17}),(\ref{eq19}) and (\ref{eq20}) are
all Fr\'{e}chet derivatives. Taking a second derivative simply
involves a single iteration of the Fr\'echet operation
\cite{BB82}. For the first derivative to exist, both the
functional and the function space must meet smoothness criteria.
The first derivative remains a functional, which for the second
derivative to exist, must remain smooth. This condition is
implicitly assumed for $v\xc[n]$ in all of DFT and TDDFT, and we
presume it here as well. We conclude that all of the second
functional derivatives encountered in TDDFT are perfectly well
defined iterations of the Fr\'{e}chet derivative operation. These
include \ben
\chi(\br,t;\br',t')=\frac{\delta^2A[v]}{\delta{v(\br',t')}{\delta{v(\br,t)}}}
\label{eq30} \een and $\chi_s$ as well as $f\xc$. All have a
retarded dependence on $t$ and $t'$ and are decidedly not
symmetric in $\br,t$ and $\br',t'$. Similarly, $\chi^{-1}$ and
$\chi_s^{-1}$ can be expressed as second derivatives, e.g. \ben
\chi^{-1}(\br,t;\br',t')=-\frac{\delta{v}(\br,t)}{\delta{n}(\br',t')}=\frac{\delta^2B[n]}{\delta{n(\br',t')}\delta{n(\br,t)}}~~,
\label{eq31} \een have retarded time dependence (Appendix A), and
are not symmetric in $\br,t$ and $\br',t'$.

We therefore agree with the main conlcusion of Amusia and Shaginyan
\cite{AS98},\cite{AS01} and Harbola and Banerjee
\cite{HB99},\cite{H01} that there is no conflict between the symmetry
and the causality. However, the way out of the dilemma is not by finding
symmetry in the inverse response functions, but by recognizing that
second-functional derivatives need {\em not} be symmetric functions of the
time variables. To
understand how this asymmetry can come about in a second functional
derivative, consider that functionals are defined on three
levels. First, there is the space on which the functions are defined;
second, there is the function space on which the functionals are
defined; and third, there is the definition of the functional. For
example, $A[v]$ is defined through Eq.(\ref{eq10}) and the map $M_1$,
on the function space $\sV$ within which the potentials $v$ are
supported on ${\mathbb R}^3\times(t_0,t_1)$. For the total functional
derivative $\delta{A}[v]/\delta{v}(\br,t)$ to exist and equal
$-n(\br,t)$, first $\sV$ must be smooth enough that variations
$\delta{v}(\br,t)$ exist which can be taken continuously to
zero. Following Lieb \cite{L82}, we have defined $\sV$ for this to be
the case. Second, the functional $A[v]$ must be smooth enough that the
resulting variation in it, $\delta{A}[v]$, exists, is linear in
$\delta{v}(\br,t)$, and goes continuously to zero with
$\delta{v}(\br,t)$. $A[v]$ meets that criterion. The functional
derivative $\delta{A}[v]/\delta{v}(\br,t)$ is then defined through
Fr\'echet's theory of linear functionals \cite{BB82}. Similarly, for
$\delta{n}(\br,t)/\delta{v}(\br',t')=\delta^2A[v]/\delta{v}(\br',t')\delta{v}(\br,t)$
to exist, $n(\br,t)$ need only meet the smoothness criterion as a
functional of $v$, which it does through the definition of $\sN_v$.

The requirement for the applicability of Schwarz's lemma, that the
second derivative be invariant with respect to interchange of the
order of differentiation, is that the first level of support, the
space on which the function is defined, be unchanged by the first
functional differentiation. That is not the case here, and
Schwarz's lemma does not apply. In $A[v]$, $v$ is supported on
${\mathbb R}^3\times(t_0,t_1)$, but in $n[v]$, the first
derivative, $v$ is supported on ${\mathbb R}^3\times(t_0,t)$
precluding the applicability of Schwarz's lemma (see also the discussion in Appendix B). If $t'>t$ in
$\delta^2A[v]/\delta{v}(\br',t')\delta{v}(\br,t)$, it must vanish,
destroying symmetry while remaining a well-defined second
functional derivative \cite{BB82}. In the Keldysh action
functional used by van Leeuwen \cite{L98,L01}, the time-ordered
contour on which the action is defined provides the support for
the time-dependence of the potential $v$. Functional
differentiation of the Keldysh action does not modify this
support, and so the second functional derivative remains symmetric
in that support. Transformation from the Keldysh time contour back
to real time introduces the asymmetry without changing the fact
that a second functional derivative was taken. Thus, van Leeuwen
has, in effect, proved that second functional derivatives need not
be symmetric. We conclude that all functional derivatives in the
RG formulation are well defined, both first and second, and that
Eq.(\ref{eq25}), a relation among second functional derivatives,
contains no inconsistencies. One thus has a choice - one can base
TDVR TDDFT on the RG action or on the Keldysh action. How to
generalize the Keldysh action so as to provide a basis for TDNR
TDDFT is not now clear. Accordingly, we choose to base our
development of TDNR TDDFT on the RG action.

\end{subsubsection}
\end{section}

\begin{section}{$N$-representable TDDFT}

The HZM construction \cite{H80} establishes that at each time $t$,
there is an infinite set of wave functions $\Psi(t)$ which yield
any preset $n(\br,t)$ in $\sN$ via the mapping
$M_2:\sP\rightarrow\sN$, \ben
n(\br,t)=\left(\Psi(t),\hat{n}(\br)\Psi(t)\right)~~, \label{eq36}
\een with $n(\br,t_0)=n_0(\br)$, Eq.(\ref{eq14}). The task in
constructing an $N$-representable TDDFT is to select a single
member of that set so that $M_2$ becomes one-to-one and
invertible, i.e. to find $M_2^{-1}:\sN\rightarrow\sP$. $M_2^{-1}$
should meet the following four criteria. 1.) It should be
universal; 2.) it should require searching only in $\sP$ and not
in $\sP$ and $\sV$; 3.) it should subsume the mapping $M_2'^{-1}$
of $v$-representable TDDFT; and 4.) it should provide a
stationarity principle.

\begin{subsubsection}*{Previous work}

Apart from formulations applicable to special classes of
potentials \cite{B81}, there are two proposals for the formulation
of NR TDDFT. That of Kohl and Dreizler \cite{KD86} does not meet
criterion 2.) and, as a consequence, cannot meet criterion 4.) as
well. That of Ghosh and Dhara \cite{GD88} does not produce
$N$-representablility, only $v$-representability. Their Theorem 4
can be restated as defining the map $M_{GD}:\sN\rightarrow\sP$,
\ben \Psi[n]=ARG\left\{{\rm
STAT}_{\Psi\rightarrow{n}}B[\Psi]\right\}~~. \label{eq37} \een
However, since in Eq.(\ref{eq37}) one searches only for a {\em
stationary} point of $B[\Psi]$ in $\Psi$, one is allowed to relax
the subsidiary condition (\ref{eq36}) by a Legendre
transformation. Eq.(\ref{eq37}) then becomes \ben
\Psi[n]=ARG\left\{{\rm
STAT}_{\Psi}\left\{B[\Psi]-\int_{t_0}^{t_1}dt\int{d\br}~\Lambda(\br,t)n(\br,t)\right\}\right\}
\label{eq38} \een Now, the Lagrange multiplier $\Lambda(\br,t)$
will exist if and only if $n(\br,t)$ is $v$-representable, in
which case Eq.(\ref{eq38}) becomes \ben \Psi[n]=ARG\left\{{\rm
STAT}_{\Psi}~A[\Psi,\Lambda]\right\} \label{eq39} \een with
$\Lambda\subset\sV$, a potential. Thus, the map defined by
Eq.(\ref{eq37}) is identical to the map
$M_2'^{-1}:\sN_v\rightarrow\sF$ defined by RG.
Eqs.(\ref{eq37})-(\ref{eq39}) should therefore be rewritten with
$\Phi[n]$ replacing $\Psi[n]$. What Ghosh and Dhara have actually
accomplished is to find a simpler and more direct proof of
$v$-representability than the original proof of RG.

\end{subsubsection}
\begin{subsubsection}*{Our approach}

We note that the stationary point $\Phi[n]$ of $B[\Psi]$ in
Eq.(\ref{eq37}) is unique in the subspace $\sF_n$
($\Psi\rightarrow{n}$) of $\sP$ for $n\subset\sN_v$. The
partial functional derivative of $A[\Psi,v]$, its gradient in $\sF_n$,
vanishes uniquely there,
\begin{eqnarray}
\nonumber
\left.\partial_{\Psi^*}A[\Psi,v]\right)_{v,n}&=&\left.\partial_{\Psi^*}B[\Psi]\right)_n=\left[i\partial_t-\hat{H}\right]\Psi\\
&=&0~~;~~n\subset\sN_v,\Psi=\Phi[n]~~.
\label{eq40}
\end{eqnarray}

Thus the magnitude squared of the gradient,
\begin{subequations}
\ben
\int_{t_0}^{t_1}dt\left(\partial_{\Psi^*}A,\partial_{\Psi^*}A\right)_{v,n}=\int_{t_0}^{t_1}dt\left(\partial_{\Psi^*}B,\partial_{\Psi^*}B\right)_n
\label{eq41a} \een has a unique minimum there as well. As the
search can be restricted to normalized $\Psi$'s without
penalty, it follows from eq.(\ref{eq41a}) that \ben
\int_{t_0}^{t_1}dt\left(\partial_{\Psi^*}A,\partial_{\Psi^*}A\right)_{v,n}=\int_{t_0}^{t_1}dt\left(\Psi,[i\partial_t-\hat{H}]^2\Psi\right)~~
\label{eq41b} \een
\end{subequations}
holds as well because of the consequent hermiticity of
$i\partial_t$.

On the other hand, no such minimum can exist in the magnitude of
the gradient for an $N$-representable $n\subset\sN$ which is not
$v$-representable. A similar situation exists in time-independent
DFT. A minimum exists in the functional
$E[\Psi]=\left(\Psi,\hat{\cal H}\Psi\right)$ for
$\Psi\rightarrow{n}$ if and only if $n$ is $v$-representable. If we suppose that a minimum exists under the constraint of fixed $n$ for $n$ not $v$-representable, the constant can be eliminated by a Legendre transformation. The Lagrange multiplier then simply adds to the external potential contradicting the hypothesis that $n$ is not $v$-representable as in the arguments associated with Eqs.(\ref{eq37})-(\ref{eq39}). For
a general $N$-representable $n$, there is only an infimum in both
$(\Psi,\hat{\cal H}\Psi)$ and $(\Psi,\hat{H}\Psi)$ at the same
point. The Levy-Lieb constrained search algorithm makes use of
this infimum to define the density functional for
$N$-representable densities:
\begin{subequations}
\ben
E[n]=INF_{\Psi\rightarrow{n}}E[\Psi]~~;
\label{eq42a}
\een
\ben
\Psi[n]=ARG\left\{INF_{\Psi\rightarrow{n}}E[\Psi]\right\}~~.
\label{eq42b}
\een
\end{subequations}

An analogous constrained search algorithm can be constructed for
TDDFT from the magnitude of the gradient \cite{LMnote}:
\begin{subequations}
\begin{eqnarray}
\nonumber
\Psi[n]&=&ARG\left\{INF_{\Psi\rightarrow{n}}\int_{t_0}^{t_1}{dt}\left(\partial_{\Psi^*}A,\partial_{\Psi^*}A\right)_{v,n}\right\}\\
&=&ARG\left\{INF_{\Psi\to{n}}\int_{t_0}^{t_1}{dt}\left(\partial_{\Psi^*}B,\partial_{\Psi^*}B\right)_n\right\}
\label{eq43a}
\\
\nonumber
&=&ARG\left\{INF_{\Psi\to{n}}\int_{t_0}^{t_1}{dt}\left(\Psi,[i\partial_t-\hat{H}]^2\Psi\right)\right\}~~;
\end{eqnarray}
\ben
A[n,v]=\int_{t_0}^{t_1}dt\left(\Psi[n],[i\partial_t-\hat{\cal H}[v]]\Psi[n]\right)~~.
\label{eq43b}
\een
\end{subequations}
We note that the quantity over which the search in
Eq.(\ref{eq41a}) is done corresponds to the time-integral of the
McLachlan functional for $\hat{H}$ \cite{M64} widely used in
formulations of semiclassical dynamics \cite{H76}. The proposed
constrained search algorithm expressed in
Eqs.(\ref{eq43a}-\ref{eq43b}) meets all of the criteria imposed
above: 1.) It is universal, not involving $v$. 2.) It requires
searching only in $\sP$. 3.) It subsumes $v$-representable $n$ for
which the infimum becomes a minimum and yields the condition \ben
\left(i\partial_t-\hat{H}\right)^2\Psi=0~~s.t.~~\Psi\to{n}~~,
\label{eq44} \een which yields the same $\Psi$ as Eq.(\ref{eq37})
or, ultimately Eq.(\ref{eq5}). Finally, 4.)~it provides a
stationarity principle since $n$ in (\ref{eq43b}) can be varied
independently of $v$, and the corresponding partial functional derivative
vanishes via Eq.(\ref{eq12}), \ben
\partial_n{A[n,v]}=0~~.
\label{eq45}
\een
Stationarity in TDDFT plays the role that minimality does in DFT regarding error reduction.

\end{subsubsection}

\end{section}

\begin{section}{Summary}

An $N$-representable time-dependent DFT has been established, and a
time-dependent analog of the Levy-Lieb constrained search algorithm
has been proposed. The central quantity in this search is the norm of
the partial functional derivative of the Runge-Gross action in the
Hilbert space of wavefunctions. The proposed constrained search meets
all of the requirements we pose: it is universal, requires searching
only in one Hilbert space, subsumes Runge-Gross $v$-representability,
and provides a stationarity principle.

\end{section}

\appendix
\begin{section}{Causality of $\chi_s^{-1}$ and $\chi^{-1}$}

As stated in Section III.3 and III.4, there is a substantial
spread of opinion in the literature with regard to the time
dependence of $\chi_s^{-1}$ and $\chi^{-1}$, differing as to
whether it is causal or symmetric. We argue here that it is
unequivocally causal, with local singularities at $t=t'^{+}$. It
is easiest to see this explicitly for the $\chi_s$ of the uniform
electron gas, which has the form $\chi_s(|\br-\br'|,t-t')$, from
space-time uniformity. Accordingly, it is diagonalized by Fourier
transforming on space and time yielding the eigenvalues
$\chi_s(\bq,\omega-i\delta)$, $\delta\downarrow 0$. The
wave-vector $\bq$ is introduced by the Fourier transform on
$\br-\br'$, the frequency $\omega$ is introduced by that on
$t-t'$, and $\delta$ is introduced by the causality of $\chi_s$,
$\chi_s(|\br-\br'|,t-t')=0$, $t>t'^{+}$.

The explicit form of $\chi_s(\bq,\omega-i\delta)$ is known
\cite{M90}. When continued to the entire complex angular frequency
plane in the process of inverting the Fourier transform on time,
its only singularities are a second-order pole at infinity and
bounded branch cuts just above the real axis. For $q<2k_F$ ($k_F$
is the Fermi wavenumber), there is one bounded branch cut at
$z=\omega+i\delta$, with $|\omega|\leq
\frac{\hbar}{2m}(2k_Fq+q^2)$; for $q>2k_F$ there are two, with
$\frac{\hbar}{2m}(-2k_Fq+q^2)\leq|\omega|\leq\frac{\hbar}{2m}(2k_Fq+q^2)$.
$\chi_s(\bq,\omega)$ has no zeros away from the branch cuts.

$\chi_s^{-1}$ is also uniform in space and time and therefore
diagonalized by Fourier transformation. It's eigenvalues are
simply \ben \chi_s^{-1}(\bq,\omega)=1/\chi_s(\bq,\omega)~~.\een
The second-order pole in $\chi_s(\bq,\omega)$ at $\omega=\infty$
yields the following behavior in $\chi_s^{-1}(\bq,\omega)$ at
$\infty$, \ben
\chi_s^{-1}(\bq,\omega)\mathop{\to}_{\stackrel{\mbox{$|\omega|\to\infty$}}{\mbox{$\delta\to
0$}}} A_s(\bq)\omega^2+B_s(\bq)+\frac{C_s(\bq)}{\omega^2}+...\een
\begin{eqnarray}
\nonumber
A_s(\bq)&=&-\frac{m}{nq^2}\\
\nonumber
B_s(\bq)&=&\frac{2}{5}\frac{E_F}{n}\left[3+\frac{5\hbar^2 q^2}{8mE_F}\right]\\
\nonumber
C_s(\bq)&=&\frac{16E_F^2
q^2}{175nm}\left[3+35\frac{\hbar^2 q^2}{8mE_F}\right]
\end{eqnarray}
where $n$ is the number of electrons per unit volume and $m$ is
the electron mass. Thus $\chi_s^{-1}$ has the form
\begin{eqnarray}
\nonumber
\chi_s^{-1}(\bq,t-t')=-A_s(\bq)\delta''((t-t')^+)+B_s(\bq)\delta((t-t')^+)\\
+\chi_s^{-1}(\bq,t-t')'~~.\label{eqA3}
\end{eqnarray}
In Eq.(\ref{eqA3}), $\delta''$ is the second derivative of the
delta function. $(\chi_s^{-1})'$ arises from the branch cut(s) and
is rigorously causal because the locations of the branch cuts in
$\chi_s^{-1}(\bq,z)$ are identical to those of $\chi_s(\bq,z)$,
being in the upper-half z-plane.

The principal change in passing from $\chi_s(\bq,\omega)$ to
$\chi(\bq,\omega)$ for the uniform electron gas is that the
free-particle excitations are replaced by quasi-particle
excitations which have finite lifetime except at $q=0$. This
causes the branch cuts to extend to infinity, but causes no change
in the formal structure of $\chi^{-1}(\bq,t-t')$ which is given by
(\ref{eqA3}) with modification of $A_s(\bq)$ to $A(\bq)$, etc.

For a non-uniform extended system for which the excitation
spectrum forms continua, be the system ordered or disordered,
there is no change in formal structure of
$\chi_s^{-1}(\br,\br';t-t')$ and $\chi^{-1}(\br,\br';t-t')$. Each
contains the local contributions $\delta''((t-t')^+)$ and
$\delta((t-t')^+)$ as well as non-local retarded contributions.
For finite systems which have at least one discrete excitation
associated with a transition from the ground state to a bound
excited state, there is a change. Each eigenvalue of
$\chi_s(\br,\br';\omega)$ or $\chi(\br,\br';\omega)$ switches from
$+\infty$ to $-\infty$ as the pole at $z=\omega+i\delta$ with
$\hbar\omega$ equal to that discrete excitation energy is crossed.
This forces the existence of a zero between discrete excitation
energies or between the highest discrete excitation energy and the
continuum threshold. Each such zero gives rise to a pole in the
corresponding eigenvalue of $\chi^{-1}(\br,\br';z)$ or
$\chi_s^{-1}(\br,\br';z)$ at the same $z$. Upon Fourier transform
to the time-domain $\chi^{-1}(\br,\br';t-t')$ and
$\chi_s^{-1}(\br,\br';t-t')$ each contains a causal contribution
from the pole which oscillates with angular frequency
corresponding to the excitation energy for $t\geq t'$ and vanishes
for $t'>t$.

In conclusion, the causality of $\chi(\br,\br';t-t')$ and
$\chi_s(\br,\br';t-t')$ forces the eigenvalues of
$\chi(\br,\br';\omega)$ and $\chi_s(\br,\br';\omega)$ to have
singularities only in the upper-half complex-frequency plane. The
corresponding eigenvalues of $\chi^{-1}(\br,\br';\omega)$ and
$\chi_s^{-1}(\br,\br';\omega)$ can therefore also have
singularities only in the upper-half plane apart from the
second-order pole at $\infty$. This arises from the fact that
$\omega$ enters $\chi$ and $\chi_s$ only in the combination
$\omega-i\delta$, $\delta\downarrow 0$, which does not change when
their eigenvalues are inverted to obtain $\chi^{-1}$ and
$\chi_s^{-1}$. Those quantities must therefore always be of the
form \ben
\hat{\chi}^{-1}(t-t')=\hat{A}\delta''((t-t')^+)+\hat{B}\delta((t-t')^+)+\hat{\chi}^{-1}(t-t')'~~,\een
where $\hat{\chi}^{-1}(t-t')'$ is nonlocal and causal in time.

\end{section}

\begin{section}{Second-functional derivative asymmetry}

In section III below Eq.(\ref{eq31}), we have pointed to the modification of the support of $v(\br,t)$ in $n[v]$ by the first functional derivative of $A[v]$ with respect to $v$ as the origin of the asymmetry of its second derivative, $\chi$. Alternatively, one can preserve the support on which the function $v(\br,t)$ is defined, but then the second level of definition, the function space on which the functional is defined, must change. Consider, for example, the map $M_1:\sV\rightarrow\sF$, Eq.(\ref{eq7}), which, together with Eq.(\ref{eq13}) defines the map $M_3:\sV\rightarrow\sN_v$, $n=n[v]$. A more explicit expression of that map would be
\ben
n=n[\hat{\cal H}[v]]~~,
\label{eqA1}
\een
according to Eq.(\ref{eq7}), in which $v(\br,t)$ is supported on $(t_0,t)$. However, Eq.(\ref{eq7}) can be rewritten as
\ben
\Phi(t)={\cal T}_L\exp\left[-i\int_{t_0}^{t_1}dt'\tilde{\cal H}_t[v]\right]\Phi_0~~,
\een
where
\begin{eqnarray}
\nonumber
\tilde{\cal H}_t[v]&=&\hat{\cal H}[v]~~,~~t'\in (t_0,t);\\
&=&0~~~~~~,~~t'\in~t,t_1)~~.
\end{eqnarray}
Thus, by changing the operator space on which the argument of the functional, now the operator $\tilde{\cal H}_t[v]$, is defined, we have formally restored the support of $v$ to $(t_0,t_1)$. However, that does not eliminate the asymmetry; it trivially shifts the location of its origin, viz
\ben
\frac{\delta n(\br,t)}{\delta v(\br',t')}=\frac{\delta n(\br,t)}{\delta\hat{\cal H}_t}\frac{\delta\hat{\cal H}_t}{\delta v(\br',t')}=0~~,~~t'>t~~.
\een
\end{section}

\begin{acknowledgements}
We are grateful to Kieron Burke for critical and stimulating discussions in response to which Appendix B was added to the paper. We thank Neepa Maitra for calling relevant references to our attention. We also thank A.K. Rajagopal for bringing his pioneering work on the symmetry-causality paradox to our attention. This research was supported in part by ONR grant N00014-01-1-0365, ONR/DARPA grant N00014-01-1-1061 and NSF grant no. CHE-9875091.
\end{acknowledgements}


\begin{thebibliography}{0}
\bibitem{22}
Van Leeuwen \cite{L01} has argued that by imposing stationarity on
$A[\Psi,v]$ with respect to the two variations $\delta\Psi$ and
$i\delta\Psi$, one can derive Eq.(\ref{eq5}) without imposing
condition (\ref{eq11}) and thus eliminate distinguishing $t_1$
from $t$ $\forall{t}\in(t_0,t_1~$. But normalization then requires
that $\left(\delta\Psi,\Psi\right)+\left(\Psi,\delta\Psi\right)=0$
and
$\left(i\delta\Psi,\Psi\right)+\left(\Psi,i\delta\Psi\right)=0$
hold simultaneously, implying that
$\left(\delta\Psi,\Phi\right)=0~\forall{t}\in(t_0,t_1)$, a much
more severe restriction than Eq.(\ref{eq11}). L\"{o}wdin and
Mukherjee \cite{LM72} had proposed this constraint earlier.
\bibitem{LMnote}
L\"{o}wdin and Mukherjee \cite{LM72} have already considered the
gradient (\ref{eq12}), but constrained only to constant $v$ and
not both $v$ and $n$ constant as in Eq.(\ref{eq40}). They used its norm
in $\sS~, (\Theta_{\Psi^*},\Theta_{\Psi^*})$, as a measure of the
error for $\Psi\neq\Phi[v]$. Minimizing that error separately for
each time leads to the Schr\"{o}dinger equation (\ref{eq5}). We
add below a constant $n$ constraint, integration over time, and
the infimum operation to generalize the LL search to the time
domain.
\end{thebibliography}
\end{document}